\overfullrule=0pt
\input harvmac

\lref\jusinskas{
   N.~Berkovits and R.~Lipinski Jusinskas,
  ``Light-Cone Analysis of the Pure Spinor Formalism for the Superstring,''
JHEP {\bf 1408}, 102 (2014).
[arXiv:1406.2290 [hep-th]].
}
\lref\AisakaUD{
  Y.~Aisaka, L.~I.~Bevilaqua and B.~C.~Vallilo,
  ``On semiclassical analysis of pure spinor superstring in an $AdS_5$ x $S^5$ background,''
JHEP {\bf 1209}, 068 (2012).
[arXiv:1206.5134 [hep-th]].
}

\lref\wittena{
E.~Witten,
``D = 10 superstring theory,''
Prog. Math. Phys. {\bf 9}, 395-408 (1983).
}

\lref\brink{
 L.~Brink, M.~B.~Green and J.~H.~Schwarz,
  ``Ten-dimensional Supersymmetric {Yang-Mills} Theory With SO(8) - Covariant Light Cone Superfields,''
Nucl.\ Phys.\ B {\bf 223}, 125 (1983)..
}

\lref\BerkovitsYR{
  N.~Berkovits and O.~Chandia,
  ``Superstring vertex operators in an AdS(5) x S**5 background'',
Nucl.\ Phys.\ B {\bf 596}, 185 (2001).
[hep-th/0009168].
}

\lref\BerkovitsRB{
  N.~Berkovits,
``Covariant quantization of the superparticle using pure spinors,''
JHEP {\bf 0109}, 016 (2001).
[hep-th/0105050].
}

\lref\GalperinAV{
  A.~Galperin, E.~Ivanov, S.~Kalitsyn, V.~Ogievetsky and E.~Sokatchev,
``Unconstrained N=2 Matter, Yang-Mills and Supergravity Theories in Harmonic Superspace,''
Class.\ Quant.\ Grav.\  {\bf 1}, 469 (1984)..
}

\lref\vallilo{
  B.~Vallilo and L.~Mazzucato,
  ``The Konishi Multilpet at Strong Coupling,''
JHEP {\bf 1112}, 029 (2011).
[arXiv:1102.1219 [hep-th]].
}

\lref\MikhailovAF{
  A.~Mikhailov,
 ``Finite dimensional vertex,''
JHEP {\bf 1112}, 005 (2011).
[arXiv:1105.2231 [hep-th]].
}

\lref\mikh{
  A.~Mikhailov and R.~Xu, ``BRST cohomology of the sum of two pure spinors,''
to appear.
}

\lref\minahan{
  J.~Minahan,
  ``Holographic three-point functions for short operators,''
JHEP {\bf 1207}, 187 (2012).
[arXiv:1206.3129 [hep-th]].
}

\lref\BerkovitsGA{
  N.~Berkovits,
  ``Simplifying and Extending the AdS(5) x S**5 Pure Spinor Formalism,''
JHEP {\bf 0909}, 051 (2009).
[arXiv:0812.5074 [hep-th]].}

\lref\BerkovitsMaz{
  N.~Berkovits and L.~Mazzucato,
`` Taming the b antighost with Ramond-Ramond flux,''
JHEP {\bf 0111},  019 (2010).
[arXic:1004.5140 [hep-th].}


\lref\BerkovitsBT{
  N.~Berkovits,
  ``Pure spinor formalism as an N=2 topological string'',
JHEP {\bf 0510}, 089 (2005).
[hep-th/0509120].}

\lref\BerkovitsXU{
  N.~Berkovits,
 ``Quantum consistency of the superstring in AdS(5) x S**5 background,''
JHEP {\bf 0503}, 041 (2005).
[hep-th/0411170].
}


\lref\BerkovitsFE{
  N.~Berkovits,
  ``Super Poincare covariant quantization of the superstring'',
JHEP {\bf 0004}, 018 (2000).
[hep-th/0001035].
}


\lref\MazzucatoJT{
  L.~Mazzucato,
  ``Superstrings in AdS'',
[arXiv:1104.2604 [hep-th]].
}


\lref\HeslopNP{
  P.~Heslop and P.~S.~Howe,
  ``Chiral superfields in IIB supergravity'',
Phys.\ Lett.\ B {\bf 502}, 259 (2001).
[hep-th/0008047].
}







\lref\SohniusWK{
  M.~F.~Sohnius,
  ``Bianchi Identities for Supersymmetric Gauge Theories,''
Nucl.\ Phys.\ B {\bf 136}, 461 (1978).
}

\lref\ArutyunovGA{
  G.~Arutyunov and S.~Frolov,
  ``Foundations of the $AdS_5 \, x \, S^5$ Superstring. Part I,''
J.\ Phys.\ A {\bf 42}, 254003 (2009).
[arXiv:0901.4937 [hep-th]].
}

\lref\MetsaevIT{
  R.~R.~Metsaev and A.~A.~Tseytlin,
  ``Type IIB superstring action in AdS(5) x S**5 background,''
Nucl.\ Phys.\ B {\bf 533}, 109 (1998).
[hep-th/9805028].
}


\lref\sen{
A.~Sen and E.~Witten,
``Filling the gaps with PCOs,''
JHEP 09, 004 (2015)
doi:10.1007/JHEP09(2015)004
[arXiv:1504.00609 [hep-th]].}

\lref\HoweSRA{
  P.~S.~Howe and P.~C.~West,
  ``The Complete N=2, D=10 Supergravity'',
Nucl.\ Phys.\ B {\bf 238}, 181 (1984).
}

\lref\BerkovitsULM{
  N.~Berkovits,
  ``Sketching a Proof of the Maldacena Conjecture at Small Radius,''
[arXiv:1903.08264 [hep-th]].
}
\lref\mafg{
  H.~Gomez and C.~R.~Mafra,
  ``The closed-string 3-loop amplitude and S-duality,''
JHEP {\bf 1310}, 217 (2013).
[arXiv:1308.6567 [hep-th]].
}
\lref\FrolovAV{
  S.~Frolov and A.~A.~Tseytlin,
  ``Semiclassical quantization of rotating superstring in AdS(5) x S**5,''
JHEP {\bf 0206}, 007 (2002).
[hep-th/0204226].
}
\lref\ChoNFN{
  M.~Cho, S.~Collier and X.~Yin,
  ``Strings in Ramond-Ramond Backgrounds from the Neveu-Schwarz-Ramond Formalism,''
[arXiv:1811.00032 [hep-th]].
}
\lref\vallilo{
O.~Chandia and B.~Vallilo,
``A superfield realization of the integrated vertex operator in an $AdS_5\times S^5$ background,''
JHEP {\bf 1710}, 178 (2017).
[arXiv:1709.00517 [hep-th]].
}

\lref\MinahanFH{
  J.~A.~Minahan,
  ``Holographic three-point functions for short operators,''
JHEP {\bf 1207}, 187 (2012).
[arXiv:1206.3129 [hep-th]].
}
\lref\MazzucatoJaT{
  L.~Mazzucato,
  ``Superstrings in AdS,''
Phys.\ Rept.\  {\bf 521}, 1 (2012).
[arXiv:1104.2604 [hep-th]].
}

\lref\BerkovitsPS{
  N.~Berkovits and T.~Fleury,
  ``Harmonic Superspace from the $AdS_5\times S^5$ Pure Spinor Formalism,''
JHEP {\bf 1303}, 022 (2013).
[arXiv:1212.3296 [hep-th]].
}
\lref\BerkovitsULM{
  N.~Berkovits,
  ``Sketching a Proof of the Maldacena Conjecture at Small Radius,''
[arXiv:1903.08264 [hep-th]].
}
\lref\BedoyaQZ{
  O.~A.~Bedoya, L.~I.~Bevilaqua, A.~Mikhailov and V.~O.~Rivelles,
  ``Notes on beta-deformations of the pure spinor superstring in AdS(5) x S(5),''
Nucl.\ Phys.\ B {\bf 848}, 155 (2011).
[arXiv:1005.0049 [hep-th]].
}

\lref\NBdynamical{
N.~Berkovits,
``Dynamical twisting and the b ghost in the pure spinor formalism,''
JHEP 06, 091 (2013)
[arXiv:1305.0693 [hep-th]].}

\lref\NBcov{
N.~Berkovits,
``Covariant Map Between Ramond-Neveu-Schwarz and Pure Spinor Formalisms for the Superstring,''
JHEP 04, 024 (2014)
[arXiv:1312.0845 [hep-th]].}

\lref\NBuntwist{
N.~Berkovits,
``Untwisting the pure spinor formalism to the RNS and twistor string in a flat and AdS$_{5} \times$ S$^{5}$ background,''
JHEP 06, 127 (2016)
[arXiv:1604.04617 [hep-th]].}

\lref\NBsusy{
N.~Berkovits,
``Manifest spacetime supersymmetry and the superstring,''
JHEP 10, 162 (2021)
[arXiv:2106.04448 [hep-th]].}

\lref\NBC{
N.~Berkovits, O. Chandia, J. Gomide and L.N.S. Martins,
``B-RNS-GSS heterotic string in curved backgrounds,'' [arXiv:2211.06899 [hep-th]].}

\lref\nekrasov{
N.~Nekrasov, private communication.}

\lref\mafrag{
H.~Gomez, C.~R.~Mafra and O.~Schlotterer,
``Two-loop superstring five-point amplitude and $S$-duality,''
Phys. Rev. D 93, no.4, 045030 (2016)
doi:10.1103/PhysRevD.93.045030
[arXiv:1504.02759 [hep-th]].}

\lref\pioline{
E.~D'Hoker, C.~R.~Mafra, B.~Pioline and O.~Schlotterer,
``Two-loop superstring five-point amplitudes. Part I. Construction via chiral splitting and pure spinors,''
JHEP 08, 135 (2020)
doi:10.1007/JHEP08(2020)135
[arXiv:2006.05270 [hep-th]].}

\lref\baulieuone{
L.~Baulieu,
``SU(5)-invariant decomposition of ten-dimensional Yang-Mills supersymmetry,''
Phys. Lett. B 698, 63-67 (2011)
[arXiv:1009.3893 [hep-th]].}

\lref\costello{K.~Costello and S.~Li,
``Quantization of open-closed BCOV theory, I,''
[arXiv:1505.06703 [hep-th]].}

\lref\NBufive{
N.~Berkovits,
``Quantization of the superstring with manifest U(5) super-Poincare invariance,''
Phys. Lett. B 457, 94-100 (1999)
[arXiv:hep-th/9902099 [hep-th]].}

\lref\NBooguri{
N.~Berkovits, H.~Ooguri and C.~Vafa,
``On the world sheet derivation of large N dualities for the superstring,''
Commun. Math. Phys. 252, 259-274 (2004)
[arXiv:hep-th/0310118 [hep-th]].}

\lref\vertical{
A.~Sen,
``Off-shell Amplitudes in Superstring Theory,''
Fortsch. Phys. 63, 149-188 (2015)
[arXiv:1408.0571 [hep-th]].}

\lref\baulieu{
L.~Baulieu,
``Transmutation of pure 2-D supergravity into topological 2-D gravity and other conformal theories,''
Phys. Lett. B 288, 59-68 (1992)
doi:10.1016/0370-2693(92)91954-8
[arXiv:hep-th/9206019 [hep-th]].}

\lref\berknek{N.~Berkovits and N.~Nekrasov,
``Multiloop superstring amplitudes from non-minimal pure spinor formalism,''
JHEP 12, 029 (2006)
[arXiv:hep-th/0609012 [hep-th]].}

\lref\topological{
N.~Berkovits,
``Pure spinor formalism as an N=2 topological string,''
JHEP 10, 089 (2005)
[arXiv:hep-th/0509120 [hep-th]].}

\lref\siegelqm{
W.~Siegel,
``Classical superstring mechanics,''
Nucl. Phys. B 263, 93-104 (1986).}

\lref\sorokin{D.~P.~Sorokin,
``Superbranes and superembeddings,''
Phys. Rept. 329, 1-101 (2000)
[arXiv:hep-th/9906142 [hep-th]].}

\def\a{{\alpha}}

\def\l{{\lambda}}
\def\lb{{\overline\lambda}}

\def\wb{{\overline w}}

\def\lb{{\overline\lambda}}
\def\b{{\beta}}

\def\g{{\gamma}}

\def\Gb{{\overline\Gamma}}

\def\e{{\epsilon}}
\def\s{{\sigma}}

\def\L{{\Lambda}}

\def\half{{1\over 2}}
\def\p{{\partial}}

\def\t{{\theta}}

\def\gt{{\widetilde \g}}

\def\bt{{\widetilde \b}}

\def\phit{{\widetilde \phi}}

\def\lb{{\overline{\lambda}}}

\Title{\vbox{\baselineskip12pt
\hbox{}}}
{{\vbox{\centerline{Equivalence Proof of Pure Spinor and }
\smallskip
\centerline{ Ramond-Neveu-Schwarz Superstring Amplitudes}}} }
\bigskip\centerline{Nathan Berkovits\foot{e-mail: nathan.berkovits@unesp.br}}
\bigskip
\centerline{\it ICTP South American Institute for Fundamental Research}
\centerline{\it Instituto de F\'\i sica Te\'orica, UNESP - Univ. 
Estadual Paulista }
\centerline{\it Rua Dr. Bento T. Ferraz 271, 01140-070, S\~ao Paulo, SP, Brasil}
\bigskip

\vskip .1in

A new manifestly spacetime-supersymmetric prescription for superstring amplitude computations is given using the pure spinor formalism which does not contain subtleties from poles in the pure spinor ghosts. This super-Poincar\'e covariant prescription is related by a U(5)-covariant field redefinition to the Ramond-Neveu-Schwarz amplitude prescription where the pure spinor parameterizes the choice of $SO(10)/U(5)$. For $F$-term scattering amplitudes which preserve a subset of the spacetime supersymmetries, the new pure spinor amplitude prescription reduces to the previous pure spinor prescription.

\vskip .1in

\Date {November 2024}
\newsec{Introduction}

Most of our current understanding of superstring theory comes from the computation of perturbative superstring scattering amplitudes, and the two main formalisms for computing these amplitudes are the Ramond-Neveu-Schwarz (RNS) formalism and the pure spinor formalism. The RNS formalism has the advantage of being based on a geometrical formulation of the worldsheet theory on N=1 super-Riemann surfaces, but has the disadvantage that spacetime supersymmetry is not manifest. On the other hand, the pure spinor formalism has manifest spacetime supersymmetry which simplifies the computation of multiloop amplitudes, but there is no obvious geometric formulation of the worldsheet theory. All amplitudes that have been computed up to now using the two formalisms coincide, but there is no general proof of their equivalence. Furthermore, there are different types of subtleties in multiloop computations using the two formalisms, and an equivalence proof would need to relate these subtleties. The RNS amplitude prescription has subtleties coming from dependence on the location of picture-changing operators, and the pure spinor prescription has subtleties coming from poles when the pure spinor ghosts vanish. 

In this paper, the two amplitude prescriptions will be proven to be equivalent by relating them (after Wick-rotation to SO(10) signature) to a U(5)-covariant prescription \NBufive\ which is a conformally invariant extension of the 
U(4)-covariant light-cone Green-Schwarz prescription.  In this U(5)-covariant prescription, the ten components of the RNS spacetime vector variable $\psi^m$ are related by a field redefinition to 5 components of the Green-Schwarz spacetime spinor variable $\theta^\alpha$ and its conjugate momentum $p_\a$. This prescription can be made SO(10)-covariant by adding bosonic variables $\l^\a$ satisfying the d=10 pure spinor constraint $\l\g^m\l =0$ which parameterize the choice of $SO(10)/U(5)$, and including the remaining 11 components of $\t^\a$ and $p_\a$ through the topological term $\oint \l^\a p_\a$ in the BRST operator so that the additional degrees of freedom do not modify the physical spectrum.

Using the field redefinition from the RNS variables, one obtains an extended pure spinor formalism which includes the original pure spinor variables $(x^m, \t^\a, p_\a, \l^\a, w_\a)$ together with a set of $(b,c)$ and $(\bt, \gt)$ worldsheet ghosts of conformal weight $(2,-1)$ and opposite statistics. The new pure spinor amplitude prescription involves picture-changing operators coming from the RNS picture-changing operators which resemble the composite $B$-ghost in the original pure spinor prescription \topological\ but contain an additional term. And the pure spinor vertex operators coming from the field redefinition of RNS vertex operators also contain an additional term related to the pure spinor antifield of ghost-number $+2$. For super-Yang-Mills states, the new vertex operator is
\eqn\vertexop{V = c e^{-\phit} \l^\a A_\a (x,\t) + c e^{-2\phit} \l^\a \l^\b A^*_{\a \b}(x,\t)}
where $\phit$ comes from the bosonic ghost  $\gt =\eta e^{\phit}$, and $\l^\a A_\a$ and $\l^\a\l^\b A^*_{\a\b}$ are the operators of ghost-number $+1$ and $+2$ in the massless cohomology of $\oint \l^\a d_\a$ \BerkovitsRB.

For $F$-term scattering amplitudes which preserve a subset of spacetime supersymmetries, one can show that the additional terms in the picture-changing operator and vertex operator do not contribute and the path integral over the $(b,c)$ and $(\bt, \gt)$ ghosts cancel each other out, so the new prescription reduces to the original pure spinor amplitude prescription \topological. For $D$-term scattering amplitudes which do not preserve any spacetime supersymmetries, the additional terms contribute and the new prescription has no subtleties coming from poles in the pure spinor ghosts, but has subtleties like the RNS formalism coming from dependence on the location of the picture-changing operators. At two loops and below, it is easy to see that the subtleties can be resolved to obtain manifestly super-Poncar\'e covariant expressions as in \pioline. So equivalence with the RNS prescription has been proven for all amplitudes which have been computed using the original pure spinor prescription, including $F$-terms at arbitrary loop such as \mafg\ and $D$-terms up to two loops. For $D$-terms at more than two loops, one needs to use the new pure spinor prescription to obtain super-Poincar\'e covariant expressions and it is hoped that an approach similar to \sen\ in the RNS formalism will be possible for treating the subtleties coming from the choice of the locations of picture-changing operators in these multiloop amplitudes. Note that unlike in the RNS formalism, these subtleties affect manifest Lorentz invariance but do not affect manifest spacetime supersymmetry, and should not complicate the proof of perturbative finiteness of multiloop amplitudes.

\newsec{U(5)-covariant field redefinition}

In light-cone gauge, the SO(8) RNS spacetime vector $\psi^J$ for $J=1$ to 8 can be related to the SO(8) Green-Schwarz spacetime spinor $\theta^A$ for $A=$ to 8 by splitting $\psi^J$ into $4_{+1}$ and $\overline 4_{-1}$ representations of U(4) as $(\psi^j, \psi_j)$ and bosonizing $\psi^j = e^{i \s_j}$ and $\psi_j = e^{-i \s_j}$ where $j=1$ to 4. In terms of $\s_j$, $\theta^A$ is described by $\theta^j = e^{i\s_j - {i\over 2} \sum_{k=1}^4 \s_k}$ and $\theta_j = e^{-i\s_j + {i\over 2} \sum_{k=1}^4 \s_k}$ where $(\theta^j, \theta_j)$ are $(4_{-1}, \overline 4_{+1})$ representations of U(4) \wittena. This U(4)-covariant field redefinition can be generalized to a U(5)-covariant field redefinition by defining \NBufive
\eqn\bosfive{\t^a = e^{\half\phi} e^{i\s_a - {i\over 2} \sum_{b=1}^5 \s_b}, \quad p_a = e^{-\half\phi} e^{-i\s_a + {i\over 2} \sum_{b=1}^5\s_b}}
for $a=1$ to 5 where the d=10 RNS vector $\psi^m$ for $m=1$ to 10 splits into $(5_1, \overline 5_{-1})$ representations of U(5) which are bosonized as $(\psi^a = e^{i\s_a}, \psi_a = e^{-i \s_a})$, the $(\b,\g)$ RNS ghosts have been fermionized as $(\b = \p \xi e^{-\phi}, \gamma = \eta e^\phi)$, and $\t^\a$ and $p_\a$ for $\a=1$ to 16 are d=10 spacetime spinors of opposite chirality which decompose under U(5) as $(1_{5\over 2}, \overline {10}_{\half}, 5_{-{3\over 2}})$ and $(\overline 5_{3\over 2}, {10}_{-\half}, 1_{-{5\over 2}})$ with $\t^a$ and $p_a$ being the $ 5_{-{3\over 2}}$ and
$\overline 5_{3\over 2}$ representations. It will also be useful to define the chiral boson 
\eqn\phitil{ \phit = {3\over 2} \phi - {i\over 2}\sum_{b=1}^5 \s_b}
which has no singular OPE's with $\t^a$ and $p_a$. In terms of $(\phit, \t^a, p_a, x^a, x_a, b, c, \eta, \xi)$, the RNS BRST operator is
\eqn\brstrns{ Q_{RNS} = \oint [ \eta e^{\phit} p_a \p x^a + \eta e^{2 \phit} (p^4)^a \p x_a + b \eta \p\eta e^{3 \phit} p^5 + c T +  b c \p c]}
where $T = p_a \p \t^a + \p x^a \p x_a + b \p c + \p (bc) + \bt\p\gt + \p (\bt\gt)$ and $(\bt = \p \xi e^{ -\phit}, \gt = \eta e^{\phit})$ are chiral bosons of conformal weight $(+2, -1)$ with $\eta$ defined to carry ghost number $+1$. Note that $Q_{RNS}$ is manifestly invariant under U(5) and under the 6 spacetime supersymmetries generated by $q_a = \oint p_a$ and $q_+ = \oint (\t^a \p x_a + b \eta e^{\phit})$ which transform as $\overline 5_{3\over 2}$ and $1_{-{5\over 2}}$ representations.

Under this field redefinition, the RNS gluon vertex operator in the $-1$ picture $V = c e^{-\phi} \psi^m a_m(x)$ maps to $V = c e^{-\phit} \t^a a_a(x) + c e^{-2\phit} (\t^4)_a a^a(x)$, which generalizes to the U(5)-covariant superfield $V= c e^{-\phit} A(x, \t) + c e^{-2\phit} A^*(x,\t)$ for the super-Yang-Mills multiplet. $QV =0$ implies that the U(5)-covariant superfields $A(x,\t)$ and $A^*(x,\t)$ satisfy
\eqn\sueprf{ \p_a \p^a A = \p_a \p^a A^* = ({\p\over{\p\t}})_a \p^a A + [({\p\over{\p\t}})^4]^a \p_a A^* =  [({\p\over{\p\t}})^3]_{abc} A =  [({\p\over{\p\t}})^2]_{[ab} \p_{c]}A =0,}
which implies the component expansion
\eqn\sole{A= \rho_+(x) + \t^a a_a(x) + \half\t^a\t^b \xi_{ab}(x), \quad A^* = ... + \half(\t^3)_{ab} \rho^{ab}(x)+ (\t^4)_a a^a(x) + (\t^5) \xi^+ (x),}
where $...$ can be gauged to zero and the 16 gluino polarizations are given by the linear combinations $(\xi^+, \xi_{ab} + \half\e_{abcde}\p^c \rho^{de}, \p^a \rho_+ -  \p_b \rho^{ab})$ with $\p_{[a} \xi_{bc]}=\p_a\xi^+ - \p^b \xi_{ab}=0$. Note that the terms in the vertex operator proportional to $\xi^+$ and $\xi_{ab}$ describe 4 on-shell components and correspond to RNS vertex operators in the $-\half$ picture, whereas the terms proportional to $\rho_+$ and $\rho^{ab}$ describe the other 4 on-shell components and correspond to RNS vertex operators in the $-{3\over 2}$ picture. Since the choice of RNS picture is not Lorentz invariant, the resulting scattering amplitude is only Lorentz covariant up to surface terms coming from changing the locations of the picture-changing operators. 

\newsec{Relation with pure spinor formalism}

Although the U(5)-covariant description of the previous section can be used to compute superstring amplitudes with 6 manifest spacetime supersymmetries, it is convenient to make all 16 supersymmetries manifest.
The first step to covariantize the BRST operator and vertex operator of \brstrns\ and \sole\ is to introduce the 11 bosonic and fermionic variables $(\l^+, \l_{ab}; \t^+, \t_{ab})$ and their conjugate momenta $(w_+, w^{ab}; p_+, p^{ab})$ by adding to the BRST operator the topological term $\oint [\l^+ p_+ +\half \l_{ab} p^{ab}]$ and performing the similarity transformation $Q \to e^R Q e^{-R}$ where $R = \oint c (w_+ \p\t^+ +\half w^{ab} \p\t_{ab})$. The resulting BRST operator is 
\eqn\brsttwo{ Q = \oint [\l^+ p_+ + \half\l_{ab} p^{ab} + \gt p_a \p x^a + \eta e^{2 \phit} (p^4)^a \p x_a + \eta \p\eta e^{3 \phit} p^5 (b + w_\a \p\t^\a) + c T +  b c \p c]}
where $T =  p_\a \p \t^\a + \p x^m \p x_m + w_\a \p\l^\a + b \p c + \p (bc) + \bt\p\gt + \p (\bt\gt)$ is now Lorentz covariant, $\l^\a = (\l^+, \l_{ab}, -(8\l^+)^{-1}\e^{abcde} \l_{bc}\l_{de})$ is a d=10 pure spinor satisfying $\l\g^m \l=0$, and $w_\a = (w_+, w^{ab}, 0)$ is its conjugate momentum in the gauge $w_a=0$.

The 6 manifest spacetime supersymmetry generators are $q_a = \oint p_a$ and $q_+ = \oint (p_+ + \p x_a \t^a + \gt (b + w_\a \p\t^\a))$, and it will be convenient to perform a 
similarity transformation by $R' = \oint \t^+ (\p x_a \t^a + \gt (b + w_\a \p\t^\a))$ so that $q_a = \oint (p_a + \t^+ \p x_a)$ and $q_+ = \oint p_+$. It will also be convenient to
shift $\phit \to \phit - \log \l^+$ and $w_+ \to w_+ +  (\l^+)^{-1}\p\phit$ so that $e^\phit$ does not transform under U(5) and carries ghost number $+1$. After this similarity transformation and shift, 
\eqn\brstthree{ Q = \oint  [\l^+ (p_+ -\t^a \p x_a) + \half\l_{ab} p^{ab} + \gt ((\l^+)^{-1}p_a (\p x^a + \t^a \p\t^+) + b + w_\a \p\t^\a) }
$$+
 \eta e^{2 \phit} (\l^+)^{-2} (p^4)^a \p x_a +  \eta \p\eta e^{3 \phit} (\l^+)^{-3}p^5 (b + w_\a \p\t^\a + (\l^+)^{-1}\p\phit \p\t^+)+ c T +  b c \p c].$$
Although $Q$ now has terms with $(\l^+)^{-1}$ dependence, one can avoid problems when $\l^+ =0$ by requiring that all terms with $(\l^+)^{-1}$ dependence are spacetime supersymmetric under $q_+ = \oint p_+$, i.e. all terms with $(\l^+)^{-1}$ dependence must be independent of the $\t^+$ zero mode. This is necessary since $Q(\t^+ (\l^+)^{-1} + c\t^+\p\t^+ (\l^+)^{-2}) =1$, so the cohomology would be trivial if terms were allowed to depend on both the $\theta^+$ zero mode and $ (\lambda^+)^{-1}$.

The next step in covariantizing the BRST operator is to perform a similarity transformation involving $\t_{ab}$ which transforms \brstthree\ into the manifestly spacetime supersymmetric expression
\eqn\brstfour{ Q = \oint  [\l^+ d_+ +\half \l_{ab} d^{ab} + \l^a d_a  + \gt ((\l^+)^{-1}d_a \Pi^a  + b + w_\a \p\t^\a) + c T +  b c \p c}
$$+
 \eta e^{2 \phit} (\l^+)^{-3}(d^4)^a (\l^+\Pi_a + \l_{ab} \Pi^b) +  \eta \p\eta e^{3 \phit} (\l^+)^{-3}d^5 (b + w_\a\p \t^\a  + (\l^+)^{-1}\p\phit\p\t^+)]$$
where $(d_a, d^{ab}, d_+)$ are the U(5) components of $d_\a = p_\a -\half \p x_m (\g^m\t)_\a -{1\over 8} (\t\g_m \p\t) (\g^m\t)_\a$, $(\Pi^a, \Pi_a)$ are the U(5) components of
$\Pi^m = \p x^m + \half\t\g^m \p\t$, and $(d_\a, \Pi^m)$ are the Green-Schwarz-Siegel variables of \siegelqm\ which commute with the 16 spacetime supersymmetry generators $q_\a = \oint [ p_\a +  \half\p x^m (\g_m\t)_\a + {1\over 24} (\t\g^m \p\t) (\g_m \t)_\a].$

Finally, as in the non-minimal pure spinor formalism, $Q$ can be made manifestly Lorentz covariant by introducing a new pure spinor variable $\lb_\a$ and a constrained fermionic spinor $r_\a$ satisfying $\lb \g^m \lb = \lb \g^m r=0$, together with their conjugate momenta $\wb^\a$ and $s^\a$, and adding the topological term $\oint \wb^\a r_\a$ to the BRST operator so that the new variables decouple from the cohomology. After including non-minimal terms in $Q$ depending on $r_\a$ so that the BRST operator remains nilpotent, one obtains the manifestly super-Poincar\'e invariant expression
\eqn\Qfinal{ Q = \oint [ \l^\a d_ \a + \wb^\a r_\a + \gt ( \Pi^m \Gb_m  -{(\l\g^{mn}r)\over{4(\l\lb)}} \Gb_m\Gb_n +  b + w_\a \p\t^\a + s^\a \p\lb_\a) + c T +  b c \p c}
$$+
{{(\l\g^{m_1 ... m_5}\l)}\over{120}} (5\eta e^{2 \phit} \Gb_{m_1} ... \Gb_{m_4} \Pi_{m_5}  +  \eta \p\eta e^{3 \phit} \Gb_{m_1} ... \Gb_{m_5} (b + s^\a\p\lb_\a + (w_\a + \p\phit {{\lb_\a}\over{(\l\lb)}}) \p\t^\a)]$$
where $\Gb_m = \half(\l\lb)^{-1}(\lb \g_m d -{1\over 8}(\l\lb)^{-1} \lb \g_{mnp} r (w \g^{np} \l))$ is defined as in \NBdynamical\ and, to simplify the expression, $w_\a$ has been gauge-fixed such that $\lb\g^m w =0$. Nilpotence of the BRST operator of \Qfinal\ can be verified by noting that $Q = e^{R_1} e^{R_2} \oint (\l^\a d_\a + \wb^\a r_ \a + b \gt) e^{-R_2} e^{-R_1}$ where
\eqn\defU{ R_1 = \oint  c ( \Pi^m \Gb_m - {{(\l\g^{mn}r)}\over{4(\l\lb)}} \Gb_m\Gb_n + w_\a \p\t^\a + s^\a\p\lb_\a + \bt\p c),}
$$ R_2 ={1\over{120}}\oint \eta e^{2\phit} (\l\g^{m_1 ... m_5}\l)  \Gb_{m_1} ... \Gb_{m_5} .$$

After including the new worldsheet variables, the massless U(5)-covariant vertex operator of \sole\ easily generalizes to the super-Poincar\'e covariant vertex operator 
\eqn\solf{V = c e^{-\phit} A + c e^{-2\phit} A^* \quad {\rm where } \quad
A = \l^\a A_\a(x,\t), \quad A^* = \l^\a \l^\b A^*_{\a\b} (x,\t),} 
and $A_\a(x,\t)$ and $A^*_{\a\b}(x,\t)$ are the super-Yang-Mills fields and antifields defined in \BerkovitsRB. The equation of motion $QV=0$ implies that $A$ and $A^*$ satisfy
\eqn\symc{  \p_m \p^m A = \p_m \p^m A^*  =  D_m \p^m A +{1\over{24}} (\l\g^{m_1 ... m_5} \l) D_{m_1} ... D_{m_4}\p_{m_5} A^* =0, }
$$ \l^\a D_\a A = \l^\a D_\a A^* =  (\l\g^{m_1 ... m_5} \l) D_{m_1} D_{m_2} D_{m_3} A =  (\l\g^{m_1 ... m_5} \l) D_{m_1} D_{m_2} \p_{m_3} A =0$$
where $D_m \equiv (\lb \g_m)^\a D_\a$ and $D_\a = {\p\over{\p\t^\a}}-\half (\g^m \t)_\a\p_m$ is the supersymmetric derivative, which implies as in \sole\ that $A$ and $A^*$ describe an on-shell gluon and gluino. 

\newsec{Equivalence of amplitude prescriptions}

After adding the non-minimal variables $(\l^+, \l_{ab}, \t^+, \t_{ab}, \lb_+, \lb^{ab}, r_+, r^{ab})$ and their conjugate momenta $(w_+, w^{ab}, p_+, p^{ab}, \wb^+, \wb_{ab}, s^+, s_{ab})$, the RNS $N$-point $g$-loop amplitude prescription is 
\eqn\rans{{\cal A} = \int d^{3g-3+N}\tau \langle {\cal N}\prod_{r=1}^N V_r(z_r) \prod_{I=1}^{3g-3+N} \int \mu_I b \prod_{j=1}^{2g-2-P} \{Q, \xi(y_j)\} \rangle}
where $P$ is the sum of the pictures of the vertex operators, 
$\{Q, \xi(y_j)\}$ are the picture-raising operators, $\mu_I$ are Beltrami differentials associated with the $3g-3+N$ Teichmuller parameters $\tau_I$, and 
\eqn\defreg{{\cal N} = \exp [ -Q (\t^\a \lb_\a + \sum_{J=1}^ g w^{(J)}_\a s^{(J)\a} )] = \exp [-\l^\a \lb_\a - \t^\a r_\a - \sum_{J=1}^g (w^{(J)}_\a \wb^{(J)\a} + d^{(J)}_\a s^{(J)\a})] }
is a BRST-invariant regulator \topological\ which is needed to integrate over the zero modes of the non-minimal variables. 

Using the BRST operator and vertex operator of \Qfinal\ and \solf, this amplitude can be computed in a manifestly spacetime supersymmetric manner where
the picture-changing operator is          
\eqn\pco{\{Q, \xi\} =c \p \xi +  e^\phit (b +  \Pi^m \Gb_m -{(\l\g^{mn}r)\over{4(\l\lb)}} \Gb_m\Gb_n +  w_\a \p\t^\a + s^\a \p\lb_\a) }
$$+{1\over{120}} (\l\g^{m_1 ... m_5}\l) (5 e^{2 \phit} \Gb_{m_1} ... \Gb_{m_4} \Pi_{m_5}  +  2 \p\eta e^{3 \phit} \Gb_{m_1} ... \Gb_{m_5} (b + (w_\a + \p \phit{{\lb_\a}\over{(\l\lb)}}) \p\t^\a + s^\a \p\lb_\a))$$
$$ +{1\over{120}}\eta \p[  (\l\g^{m_1 ... m_5}\l) e^{3 \phit} \Gb_{m_1} ... \Gb_{m_5} (b +(w_\a + \p\phit{{\lb_\a}\over{(\l\lb)}}) \p\t^\a + s^\a \p\lb_\a)].$$
One might worry that, as in the original pure spinor amplitude prescription, the factors of $(\l\lb)^{-1}$ in $\Gb_m$ in the picture-changing operators might make
the functional integral ill-defined near $(\l\lb)=0$. Note that each factor of $(\l\lb)^{-1}$ in $\Gb_m$ is accompanied by either a factor of $\lb_\a$ or $r_\a$. So before including the factors of $r_\a$ coming from the regulator ${\cal N}$, all poles of order $(\lb)^{-n}$ are multiplied by $(r)^n$. 

However, Lorentz invariance of ${\cal A}$ in the RNS amplitude prescription implies
that the integrand of ${\cal A}$ in \rans\ is independent of $\lb_\a$ up to possible surface terms. And because of the term $\int \wb^\a r_\a$ in the BRST operator, independence of $\lb$ up to surface terms implies by BRST invariance that all factors of $r$ in the integrand of ${\cal A}$ must be proportional to surface terms. So if surface terms can be ignored, there are no factors of $r$ in the integrand of ${\cal A}$ which implies there are no poles when $(\l\lb)\to 0$. The property that ${\cal A}$ in the pure spinor amplitude prescription is Lorentz invariant up to possible surface terms is analogous to the property that ${\cal A}$ in the RNS amplitude prescription is spacetime supersymmetric up to possible surface terms

To avoid problems when $(\l\lb)\to 0$, the new pure spinor amplitude prescription of \rans\ must differ in 3 important ways from the original pure spinor amplitude prescription:
the worldsheet variables include new ``ghost'' variables $(\bt,\gt)$ and $(b,c)$ of conformal weight $(2,-1)$; the super-Yang-Mills vertex operator depends on
both the pure spinor superfields $A=\l^\a A_\a(x,\t)$ and $A^*= \l^\a \l^\b A^*_{\a\b}(x, \t)$; and the composite $B$ ghost is replaced by insertion of the picture-changing
operator of \pco. Nevertheless, it will now be shown that the new prescription coincides with the original pure spinor prescription for $F$-term amplitudes in which at least one spacetime supersymmetry is preserved, i.e. when at least one of the 16 $\t^\a$ zero modes in the path integral comes from the regulator ${\cal N}$ and not from the vertex operators $\prod_{r=1}^N V_r$. To show this,
first perform the similarity transformation by $R_1$ and $R_2$ of \defU\ which maps $Q\to Q' = \oint (\l^\a d_\a + \wb^\a r_\a + b\gt)$. After the similarity transformation,
\eqn\after{
\int\mu b \to \int\mu (b+B + \bt \p c + \p(\bt c)) \quad {\rm and} \quad \{Q, \xi\} \to b e^\phi + \{Q', {1\over{120}}e^{2\phit} (\l\g^{m_1 ... m_5}\l)\Gb_{m_1} ... \Gb_{m_5}\}}
in the amplitude prescription of 
\rans\ where $B = \Pi^m \Gb_m- {(\l\g^{mn}r)\over{4(\l\lb)}} \Gb_m\Gb_n +  w_\a \p\t^\a + s^\a \p\lb_\a$ is the composite $B$ ghost \NBdynamical\ in the original pure spinor formalism.

Naively, the term $\{Q', {1\over{120}}e^{2\phit} (\l\g^{m_1 ... m_5}\l)\Gb_{m_1} ... \Gb_{m_5}\}$ in the picture-changing operator and the term $c e^{-2\phit} \l^\a \l^\b A^*_{\a\b}(x,\t) =[Q', c\p c \p\xi e^{-3\phit} \l^\a \l^\b A^*_{\a\b}(x,\t)]$ in the vertex operator can be dropped since they are BRST-trivial. However, one needs to be careful since BRST-trivial quantities $\{Q', \L\}$ where $\L$ depends on $\lb_\a$ and all 16 $\t^\a$ zero modes may not decouple. For example, $\L = (\l^3 \t^5) (\l\lb + r\t)^{-1} \lb_\a \t^\a = ... + (\l^3 \lb r^{10})(\l\lb)^{-11} (\t)^{16}$ satisfies $Q'(\L) = (\l^3 \t^5)$ where $(\l^3\t^5)$ denotes $(\t\g_{mnp}\t)(\l\g^m\t)(\l\g^n \t)(\l\g^p\t)$. But $\langle {\cal N} (\l^3\t^5) \rangle$ is non-zero, which implies that $Q'(\L)$ does not decouple. Note that terms which explicitly depend on $\lb_\a$ are not Lorentz invariant in terms of the RNS variables and can have poles when $(\l\lb)\to 0$. 

Fortunately, the above BRST-trivial terms in the picture-changing operator and vertex operator decouple for $F$-term amplitudes since $\L$ cannot depend on all 16 $\t^\a$ zero modes if at least one of the $\t^\a$ zero modes must come from ${\cal N}$. As discussed in \berknek, any $\t^\a$ zero mode coming from ${\cal N}$ of 
\defreg\ is accompanied by $r_\a$, which implies that $F$-term amplitudes cannot have terms proportional to $(\l\lb)^{-11}$ which would make the functional integral ill-defined. After dropping the above BRST-trivial terms, the functional integral over the $(b,c)$ and $(\bt,\gt)$ ghosts cancel each other out and one obtains the usual pure spinor amplitude prescription ${\cal A} = \int d^{3g-3+N}\tau\langle {\cal N}\prod_{r=1}^N \l^\a A_{r\a}(z_r) \prod_{I=1}^{3g-3+N} \int \mu_I B   \rangle.$ So for $F$-term amplitudes, the RNS prescription has been proven to agree with the original pure spinor amplitude prescription.

\vskip 10pt
{\bf Acknowledgements:}
I would like to thank 
CNPq grant 311434/2020-7
and FAPESP grants 2016/01343-7, 2021/14335-0, 2019/21281-4 and 2019/24277-8 for partial financial support.

\listrefs

\end